# Digital Domain Power Division Multiplexed Dual Polarization Coherent Optical OFDM Transmission

Qiong Wu[1], Zhenhua Feng[1], Ming Tang[1*], Xiang Li[2], Ming Luo[2], Huibin Zhou[1], Songnian Fu[1], and Deming Liu[1]

[1] Wuhan National Lab for Optoelectronics (WNLO) & National Engineering Laboratory for Next Generation Internet Access System, School of Optical and Electronic Information, Huazhong University of Science and Technology, Wuhan, 430074, China

[2] State Key Laboratory of Optical Communication Technologies and Networks, Wuhan Research Institute of Post and Telecommunication, Wuhan 430074, Hubei, China

* tangming@mail.hust.edu.cn

**ABSTRACT**

Capacity is the eternal pursuit for communication systems due to the overwhelming demand of bandwidth hungry applications. As the backbone infrastructure of modern communication networks, the optical fiber transmission system undergoes a significant capacity growth over decades by exploiting available physical dimensions (time, frequency, quadrature, polarization and space) of the optical carrier for multiplexing. For each dimension, stringent orthogonality must be guaranteed for perfect separation of independent multiplexed signals. To catch up with the ever-increasing capacity requirement, it is therefore interesting and important to develop new multiplexing methodologies relaxing the orthogonal constraint thus achieving better spectral efficiency and more flexibility of frequency reuse. Inspired by the idea of non-orthogonal multiple access (NOMA) scheme, here we propose a digital domain power division multiplexed (PDM) transmission technology which is fully compatible with current dual polarization (DP) coherent optical communication system. The coherent optical orthogonal frequency division multiplexing (CO-OFDM) modulation has been employed owing to its great superiority on high spectral efficiency, flexible coding, ease of channel estimation and robustness against fiber dispersion. And a PDM-DP-CO-OFDM has been theoretically and experimentally demonstrated with 100Gb/s wavelength division multiplexing (WDM) transmission over 1440km standard single mode fibers (SSMFs). Two baseband quadrature phase shift keying (QPSK) OFDM signals are overlaid together with different power levels. After IQ modulation, polarization multiplexing and long distance fiber transmission, the PDM-DP-CO-OFDM signal has been successfully recovered in the typical polarization diversity coherent receiver by successive interference cancellation (SIC) algorithm. Non-orthogonal overlaid signals different in power double the system spectral efficiency and it enables flexible provisioning of quality of service (QoS) by properly adjusting power ratios of non-orthogonal multiplexed branches.

## Introduction

To meet the increasing demand of high capacity optical fiber transmission network, five available physical dimensions including time, frequency, quadrature, polarization and space have been utilized for modulation and multiplexing in optical communications[1]. In these schemes, stringent orthogonality must be satisfied to avoid inference or crosstalk from other channels so that signals can be separated individually without degrading each other's detection performance. However, one of the major problems in these orthogonal multiplexing systems such as orthogonal frequency division multiplexing (OFDM) or Nyquist WDM is that they do not allow frequency reuse of two independent signals within the same physical dimension. Recently, the non-orthogonal multiple access (NOMA), also known as power domain multiple access, is proposed as a potential candidate for the upcoming 5G wireless communication standard due to its superior spectral efficiency[2]. In NOMA, multiple users are multiplexed with different power levels using superposition coding at the transmitter side and successive interference cancellation (SIC) based multi-user detection algorithms at the receivers. The non-orthogonal feature significantly improves the capacity and throughput in both wireless and visible light communication systems by allocating the entire bandwidth to different users simultaneously[3,4]. Inspired by the idea of NOMA, we have recently proved the feasibility of using a new multiplexing dimension to enhance the capacity of direct detection optical OFDM (DDO-OFDM) system by multiplexing the spectrally overlaid signals with different power levels in digital domain[5,6].

In this paper, we further develop the idea of digital domain power division multiplexing (PDM) into dual polarization coherent optical OFDM (DP-CO-OFDM) transmission systems, which is promising for ultra-large capacity long-haul coherent optical fiber communications. After theoretical analysis and numerical simulations, we experimentally demonstrated a 100 Gb/s WDM transmission over a standard single mode fiber (SSMF) link of 1440km. The results show that system capacity can be nearly doubled when two baseband quadrature phase shift keying (QPSK) OFDM signals are overlaid together with proper power ratio before optical modulation process and decoded after phase recovery process using carefully designed SIC algorithm, which is proved to show better performance than conventional hierarchical de-mapping approach in our experiment. Besides the capacity upgrade, the PDM scheme is also flexible to allocate customized quality of service (QoS) for different subscribers by properly setting suitable power levels of non-orthogonal multiplexed branches.

## Results and discussion

### Theory

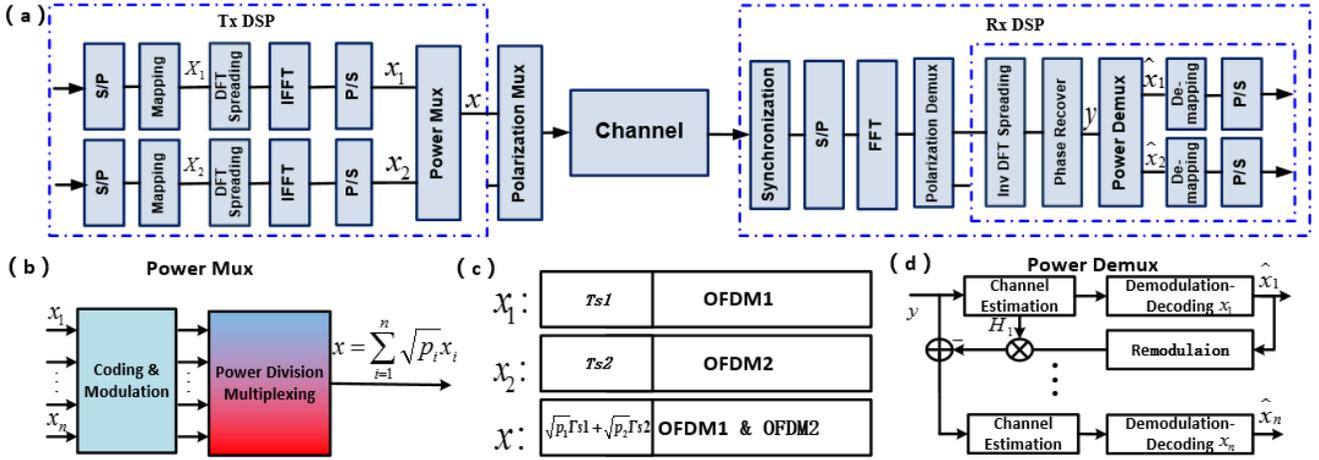

**Fig 1.** Schematic of our proposed digital domain power division multiplexed dual polarization coherent optical OFDM transmission system: (a) system model, (b) digital domain power division multiplexing module, (c) OFDM frame structure, (d) SIC algorithm based power division de-multiplexing module.

Fig 1(a) gives a global view of the system model of our proposed digital domain power division multiplexed dual polarization coherent optical OFDM (PDM-DP-CO-OFDM) transmission system. The digital signal processing (DSP) module at the transmitter side of conventional OFDM modulation includes serial-to-parallel (S/P) conversion, symbol mapping, discrete Fourier transform (DFT) spreading, inverse fast Fourier transform (IFFT), and the parallel-to-serial (P/S) conversion. Two branches of OFDM baseband signal are overlaid together with different power levels before dual polarization modulation process. After transmission along the optical fiber channel, at the receiver side, the synchronization including frame symbol aliasing and frequency offset estimation/compensation is conducted first. Then, serial-to-parallel (S/P) conversion is completed before the fast Fourier transform (FFT), after which multiple input multiple output (MIMO) processing is performed to separate signals with different polarization states. Finally, inverse DFT spreading and phase recovery are carried out before de-multiplexing overlaid signals with different power levels.

Suppose $D$ and $F$ are the DFT spreading matrix and FFT matrix respectively, $(X_i)$ and $(x_i)$ are the baseband QAM symbols and OFDM symbols respectively, then the DFT-spread OFDM modulation process could be expressed as

$$x_i = F^{-1} Z_P D X_i \ (i=1,2,...,n) \tag{1}$$

Where $Z_P$ is the zero-padding matrix used to pad the data-absent subcarriers with zeros. As shown in Fig 1(b), multiple baseband OFDM signals $(x_i)$ are linearly combined together after coding and modulation using power division multiplexing in digital domain to form one new baseband version

$$x = \sum_{i=1}^{n} \sqrt{p_i} x_i$$

$$= \sum_{i=1}^{n} \sqrt{p_i} F^{-1} Z_P D X_i$$

$$= F^{-1} Z_P D \sum_{i=1}^{n} \sqrt{p_i} X_i \qquad (2)$$

where $p_i$ is the power of $i^{th}$ signal, and the total power is normalized, namely $\sum_{i=1}^{n} p_i = 1$. The idea of our non-orthogonal scheme is to utilize power division multiplexing in terms of using the same frequency but exploiting different power levels. Therefore, the multiplexed signals should have different power levels. Without loss of generality, we assume that the power levels $(p_i)$ are arranged in descending order, namely $p_1 > p_2 > \cdots > p_n$. The OFDM frame structure is depicted in Fig 1(c) to show how two baseband OFDM signals are overlaid together. After transmission through the communication channel, suppose $h$ and $H$ are the temporal and frequency response of the baseband signal respectively, $\Lambda$ is the phase rotation matrix caused by phase noise, then the received power multiplexed signal can be expressed as

$$r = h * x \Lambda + N$$
$$= F^{-1} H F x \Lambda + N$$
$$= F^{-1} H Z_P D \sum_{i=1}^{n} \sqrt{p_i} X_i \Lambda + N \qquad (3)$$

where $N$ is the additive white Gaussian noise (AWGN). Then the received signal after OFDM demodulation, can be expressed as

$$y = D^{-1} Z_R (H^T H)^{-1} H^T F r \Lambda^{-1}$$
$$= D^{-1} Z_R (H^T H)^{-1} H^T F (F^{-1} H Z_P D \sum_{i=1}^{n} \sqrt{p_i} X_i \Lambda + N) \Lambda^{-1}$$
$$= \sum_{i=1}^{n} \sqrt{p_i} X_i + D^{-1} Z_R (H^T H)^{-1} H^T F N \Lambda^{-1}$$
$$= \sum_{i=1}^{n} H_i X_i + k N \Lambda^{-1} \qquad (4)$$

Where $Z_R$ is the zero-removing matrix used to remove the data-absent subcarriers, $H_i$ is the channel response of the $i^{th}$ multiplexed baseband QAM symbols, $k$ is a constant noise coefficient. Eventually, the detected signal will be demodulated and decoded one by one according to the descending order of power levels using SIC algorithm, whose flow chart is shown in Fig 1(d). It is worth mentioning that the overlaid signals can also be separated by hierarchical de-mapping[7].

**Simulation**

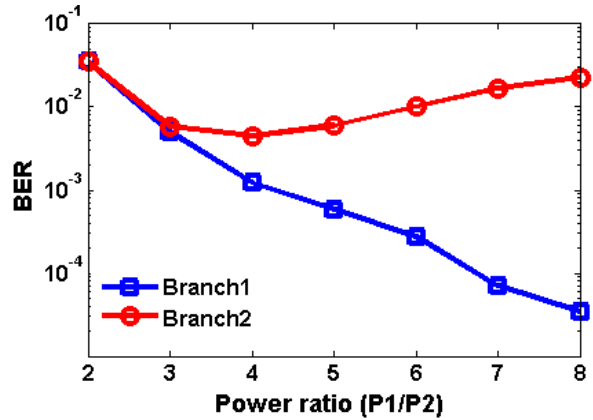

**Fig 2.** BER of both QPSK branches versus power division ratio

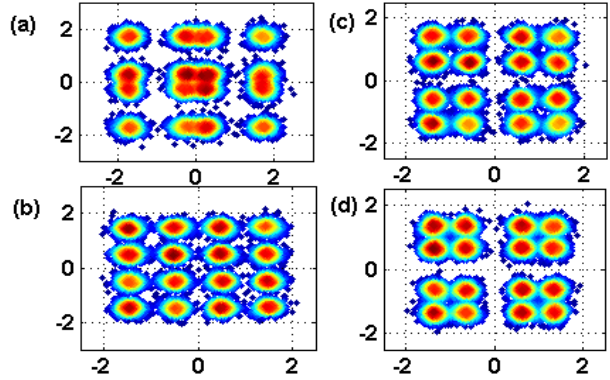

**Fig 3.** Constellation diagram: (a) 2:1, (b) 4:1, (c) 6:1, (d) 8:1

Without loss of generality, we study the digital domain power division multiplexing scheme in the scenario of two branches of QPSK-OFDM signals in this work. To investigate the relationship between the power division ratio (PDR, defined as P1/P2) and BER performance of each power multiplexed branch, we firstly conduct Matlab based simulation under the condition of AWGN channel with SNR= 8 dB. Two baseband QPSK-OFDM signals are generated from PRBS $2^{15}$ using 512 points IFFT among which 232 subcarriers are used to carry data information. From Fig 2, we can conclude that the optimized PDR to minimize the BER of the second branch with lower power is 4:1, in which case the two superposed QPSK could be emulated by a 16QAM constellation. We also show the evolution of the constellation diagram vs different PDR in Fig 3.

**Experiment**

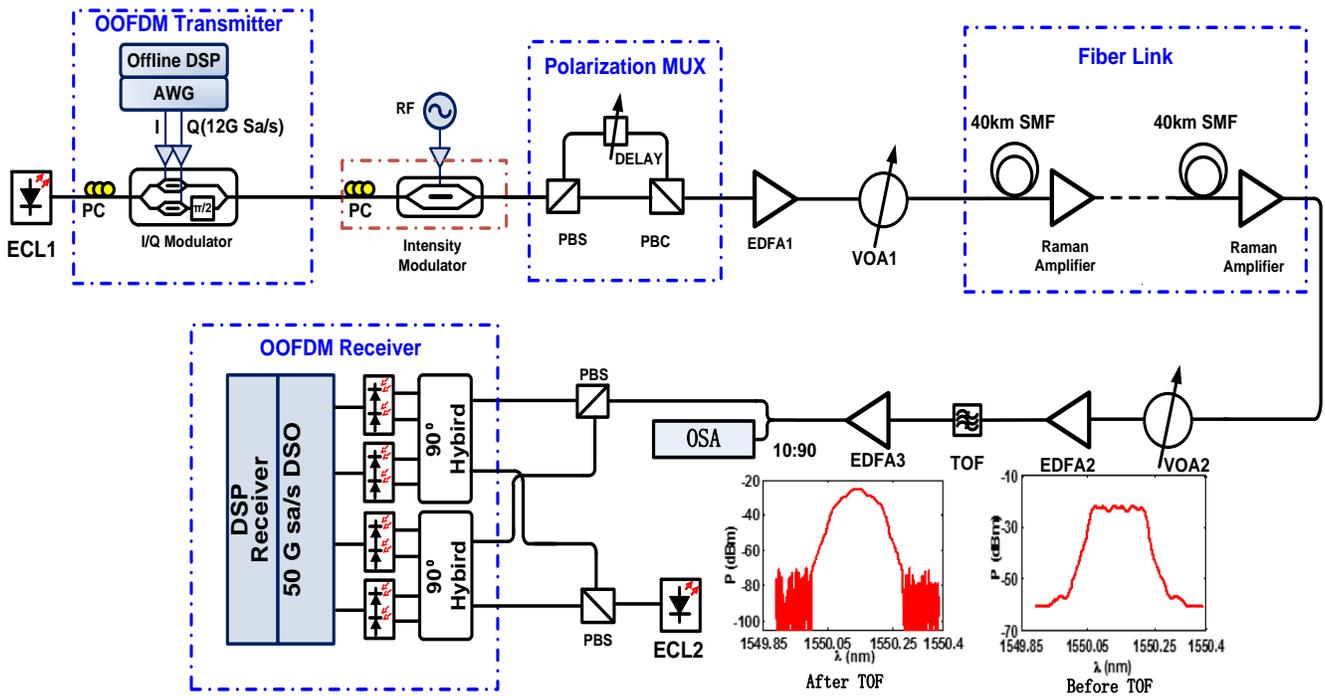

**Fig 4.** Experimental setup. ECL: External cavity laser. AWG: arbitrary waveform generator. PC: Polarization controller. PBS/C: Polarization beam splitter/combiner. EDFA: Erbium doped fiber amplifier. VOA: Variable optical attenuator. TOF: Tunable optical filter. OSA: Optical spectrum analyzer.

To verify the proposed digital domain power division multiplexing scheme, we performed further experiments using the setup shown in Fig. 4. At the transmitter, two baseband QPSK-OFDM signals are generated in Matlab originated from PRBS $2^{15}$ -1 using 512 points IFFT among which 232 subcarriers are used to carry data information. To combat the fiber dispersion, the cyclic prefix (CP) is set to 64. Each OFDM frame has 140 OFDM symbols and at the beginning of each OFDM frame are 15 training symbols (TSs), one of which is employed for OFDM timing synchronization and the rest fourteen are used for polarization de-multiplexing and channel estimation. Two branches of QPSK-OFDM signals with power division ratio 4:1 are overlaid together in digital domain and then D/A converted by an AWG (Tektronix, 7122C) at a sampling rate of 12GSa/s. After up-conversion the modulated optical signal is then duplicated to three copies by an intensity modulator in order to ensure orthogonal band multiplexing with channel spacing of 7.03125GHz[8]. Therefore, the total data rate is 3× 2×2×2×12×232/576×125/140 = 103.57 Gb/s after sub-band multiplexing. Afterward, the multi-band optical signal is polarization multiplexed by a pair of polarization beam splitter/combiner with one branch delayed by one OFDM symbol. The transmission link contains several spans of 40km SSMF whose loss is fully compensated by Raman amplifiers. At the receiver, a VOA followed by an EDFA is used to adjust the optical signal noise ratio (OSNR) of the total optical link, which can be measured by optical spectrum analyzer. A tunable optical filter (TOF) is used to select the required band before the dual polarization coherent receiver, after which two IQ components are acquired by an oscilloscope at 50GSa/s and the spectrally overlaid OFDM signal is power de-multiplexed by a SCI receiver after phase recovery of both polarization tributaries during the offline DSP.

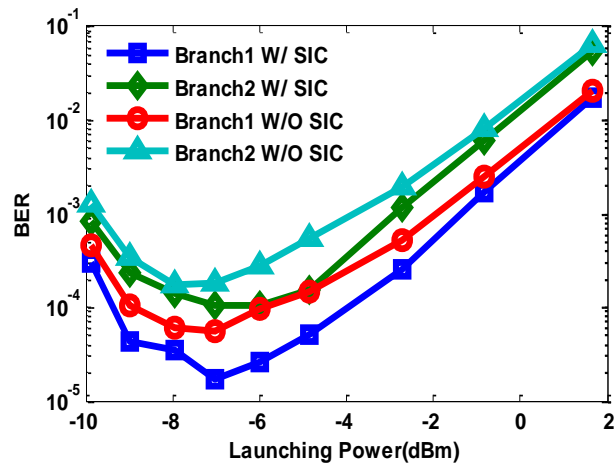

**Fig 5.** BER versus launch power with 480km transmission in both power division multiplexed branches using SIC algorithm and hierarchical de-mapping.

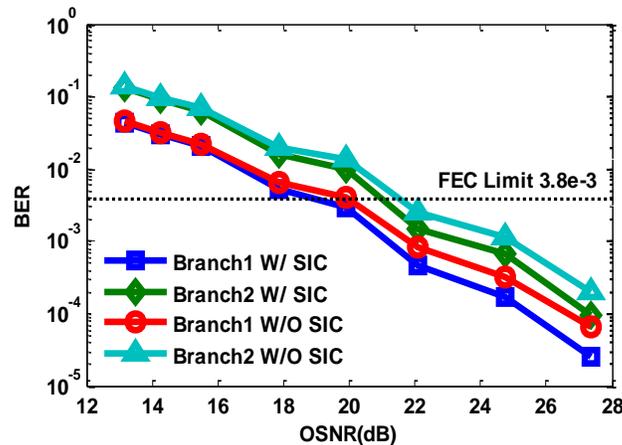

**Fig 6.** BER versus OSNR with 480km transmission in both power division multiplexed branches using SIC algorithm and hierarchical de-mapping.

In our experiment, hierarchical de-mapping is also employed to demodulate the two multiplexed QPSK branches and its performance is compared with that demodulated by SIC algorithm. In convenience, W/O SIC is used to represent hierarchical de-mapping in the following figures.

We first investigated the transmission performance of our proposed system with fiber link of 480km. As shown in Fig 5, the optimal launching power is found to be -7dBm, regardless of the power division de-multiplexing method we use. Then we vary OSNR of the received optical signal to check the BER performance of both QPSK branches with launching power fixed at -7dBm. As can be seen from Fig 6, the QPSK Branch1 allocated with higher power enjoys an OSNR redundancy of about 2dB than the QPSK Branch2 to reach the FEC limit of BER=3.8e-3, while SIC algorithm shows an OSNR improvement of about 0.5dB compared with hierarchical de-mapping.

We also confirmed the robustness of our proposed system with extended fiber links in Fig 7, which proves the feasibility of digital domain power division multiplexing in long-reach optical coherent transmission systems compatible with current wavelength division multiplexing (WDM) and space division multiplexing (SDM) technologies. Obviously, SIC algorithm can improve the transmission distance of both multiplexed QPSK branches, while the QPSK Branch1 with higher power level enjoys a transmission redundancy of about 250km than QPSK Branch2 with lower power level to reach the same FEC limit of BER=3.8e-3. To ensure identical BER performance of the two power multiplexed signals, stronger FEC coding such as soft-decision FEC can be used for weaker signals. The constellation diagrams after 480km SSMF transmission are depicted in Fig 8, which agrees well with the simulation predictions.

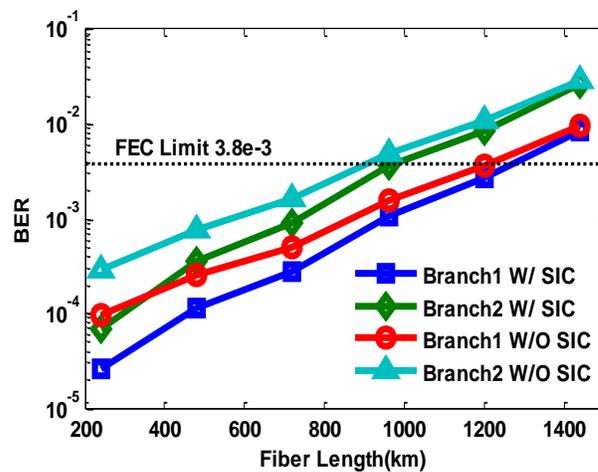

**Fig 7.** BER versus fiber length in both power division multiplexed branches using SIC algorithm and hierarchical de-mapping.

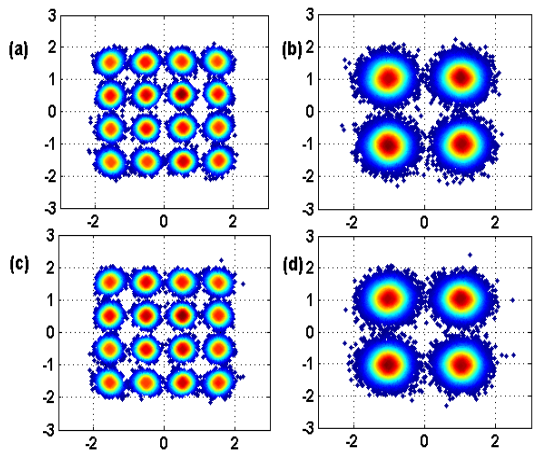

**Fig 8.** Constellation diagrams at length=480km. (a) Branch1 in Pol-X, (b) Branch2 in Pol-X, (c) Branch1 in Pol-Y, (d) Branch2 in Pol-Y.

## Conclusion

While chasing high capacity to meet the increasing traffic demands, the next generation optical network features the flexibility. Traditional physical multiplexing dimensions like time, frequency and space have little room to adjust multiplexed signal quality due to the physical stringent orthogonal condition thus sacrifice the flexibility. In this paper, we propose a DP-CO-OFDM transmission scheme based on digital domain non-orthogonal power division multiplexing, and verified it in a 100-Gb/s transmission over 1440km fiber link. Our scheme proves to be feasible with a nearly doubled system capacity when two baseband QPSK-OFDM signals of different power levels are overlaid together before optical modulation process and successively decoded after phase recovery using either SIC algorithm or hierarchical de-mapping, while SIC algorithm turns to be a better choice to improve BER performance. Besides, non-orthogonal digital domain power division multiplexing is compatible with current dual polarization WDM and SDM systems. Moreover, classified quality of service (QoS) can also be realized anytime by simply adjusting the power ratios and modulation formats of different subscribers with various bandwidth requirements and transmission distances. Above all, it is a promising technique to improve the system capacity and flexibility simultaneously for future network construction, expansion, maintenance and operation.

## Methods

**Power de-multiplexing using SIC Algorithm** As shown in Fig 1(c), The SIC process mainly consists of three steps: (1) estimate the channel response and demodulate the stronger signal $x_i$ while treating all the other signals $x_{i+1} \cdots x_n$ as interference noise; (2) re-modulate the estimated signal $\hat{x}_i$ and multiply it by the channel response $H_i$ before subtract the product from the received signal $y$ and then decode the weaker signal $x_{i+1}$; (3) continue with the aforementioned steps until all the signals are decoded. As SIC algorithm suppresses interference from stronger (earlier decoded) signals for relatively weaker (yet to be demodulated) signals, even the signals with lower power can be correctly recovered.

**Hierarchical de-mapping of two overlaid QPSK branches** Fig 9 draws the mapping constellation of single QPSK branch and two overlaid QPSK branches. When demodulate with SIC algorithm, two overlaid QPSK branches are separated first and then both de-mapped using single QPSK mapping constellation on the left. However, when demodulate with hierarchical de-mapping, two overlaid QPSK branches are straightly de-mapped using the hierarchical mapping constellation on the right, then the first two bits are allocated to the QPSK branch with higher power level while the rest two bits are allocated to the lower one.

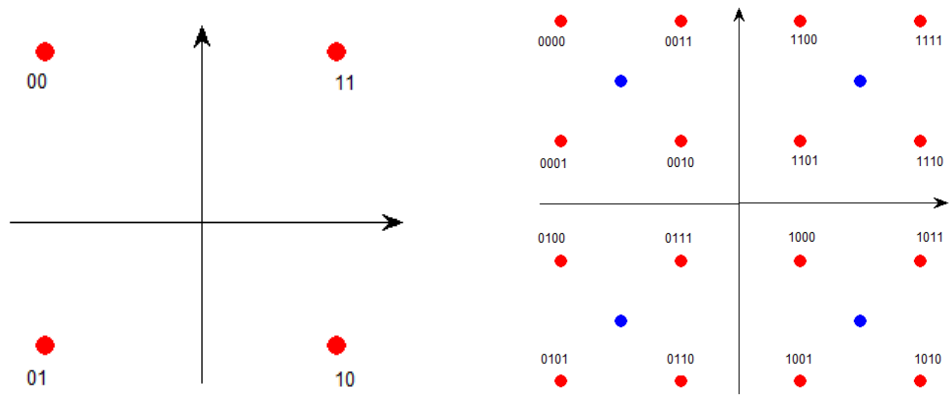

**Fig 9.** Mapping constellation of single QPSK branch and two overlaid QPSK branches

**Acknowledgements (not compulsory)**

This work was supported by the National Natural Science Foundation of China (NSFC) under Grant No. 61331010, the 863 High Technology Plan of China (2013AA013402) and the Program for New Century Excellent Talents in University (NCET-13-0235).

**Author contributions statement**

M.T., Q.W., and Z.F. conceived the study. X.L., M.L., H.Z. and Q.W. designed and did the experiment. Q.W., Z.F. and M.T. developed the theory and analyzed the data. Q.W., Z.F., M.T., S.F. and D.L. prepared the manuscript.

**Competing financial interests:** The authors declare no competing financial interests.